\documentclass[a4paper,11pt]{article}
\usepackage{pos}

\newcommand{\snn}[1]{$\sqrt{s_{_{NN}}}$ = #1\,GeV}

\title{Thermal photon measurements at PHENIX}

\author*[a]{Roli Esha (for the PHENIX Collaboration)}

\affiliation[a]{Department of Physics and Astronomy, Stony Brook University, \\
Stony Brook, NY 11790 USA.}

\emailAdd{roli.esha@stonybrook.edu}

\abstract{Photons are emitted at all stages of relativistic heavy-ion collisions and do not interact with the medium strongly. With access to the versatility of RHIC, measurements of low momentum direct photons are made possible across different system size and beam energies. An excess of direct photons, above prompt photon production from hard scattering processes, is observed for a system size corresponding to $dN_{ch}/d\eta$ of 20-30, with a large azimuthal anisotropy and a characteristic dependence on collision centrality. After subtracting the prompt photon component, the inverse slope of the spectrum is continuously increasing with the effective temperature ranging from 250 MeV/c at $p_{T}$ of 1-2 GeV/c to about 400 MeV/c at 2-4 GeV/c. Within the experimental uncertainty, there is no indication of a system size dependence of the inverse slope. In this proceeding, results from Au+Au collisions from the PHENIX experiment will be presented.}

\FullConference{HardProbes2023\\
 26-31 March 2023\\
 Aschaffenburg, Germany\\}

\begin{document}
\maketitle

\section{Introduction}
By virtue of being color neutral, photons provide snapshots of the space-time evolution of the hot and dense medium produced in relativistic heavy-ion collisions. Direct photons, defined as those which do not come from hadronic decays, are sensitive to the temperature of the medium and its measurement helps constrain initial conditions, and, sources of photon production and their emission rates.

All photon sources can be classified into: decay photons, which make up approximately 80-90\% of the total photon yield, and direct photons. Direct photons, in turn, can be further categorized into two subcategories: prompt and non-prompt photons. Prompt photons originate from sources akin to those found in $p$+$p$ collisions and their yield scales with the number of binary collisions. In addition to the well-known thermal sources originating from the Hadron Gas and the Quark-Gluon Plasma (QGP) phase, other sources that contribute to non-prompt direct photons include interactions between jets and the surrounding medium, as well as emissions occurring during the pre-equilibrium state. As the system evolves, it undergoes expansion and cooling. Consequently, earlier phases are characterized by higher temperatures and are more likely to dominate the emissions at higher transverse momentum ($p_T$).

The wealth of data and an optimized detector configuration has enabled PHENIX to measure direct photons across 7 systems and 3 collision energies over a large $p_T$ range. To ensure robustness and accuracy, three distinct methods have been employed: the calorimeter method, the virtual photon method, and the external conversion method. In this proceeding, results from the external conversion method used for analyzing the direct photons from the years 2010~\cite{PHENIX:2022qfp} and 2014~\cite{PHENIX:2022rsx} will be discussed.

\section{Direct photons}

\begin{figure}
\centerline{%
\includegraphics[height=5.45cm]{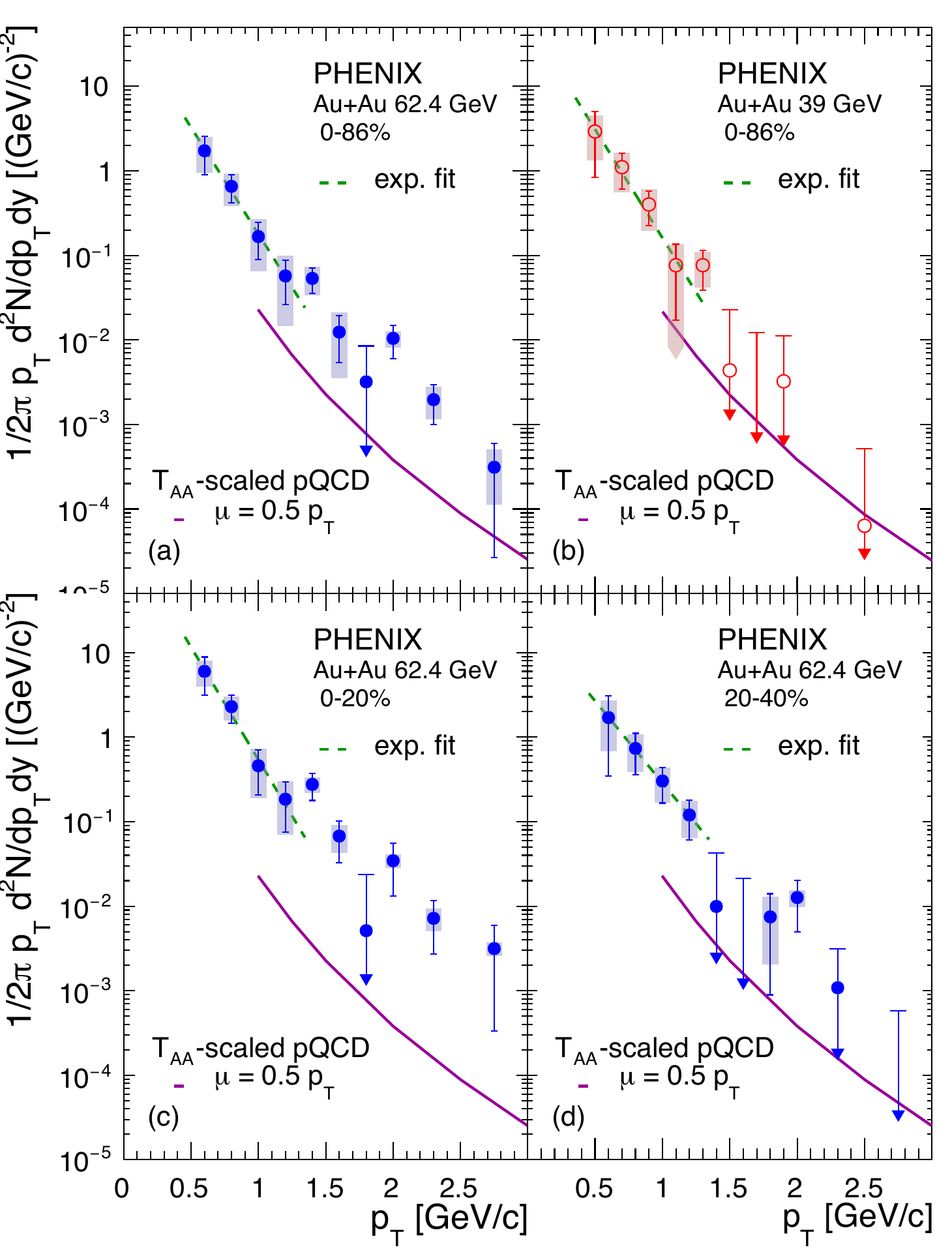} 
 \hspace{2cm}
\includegraphics[height=5.4cm]{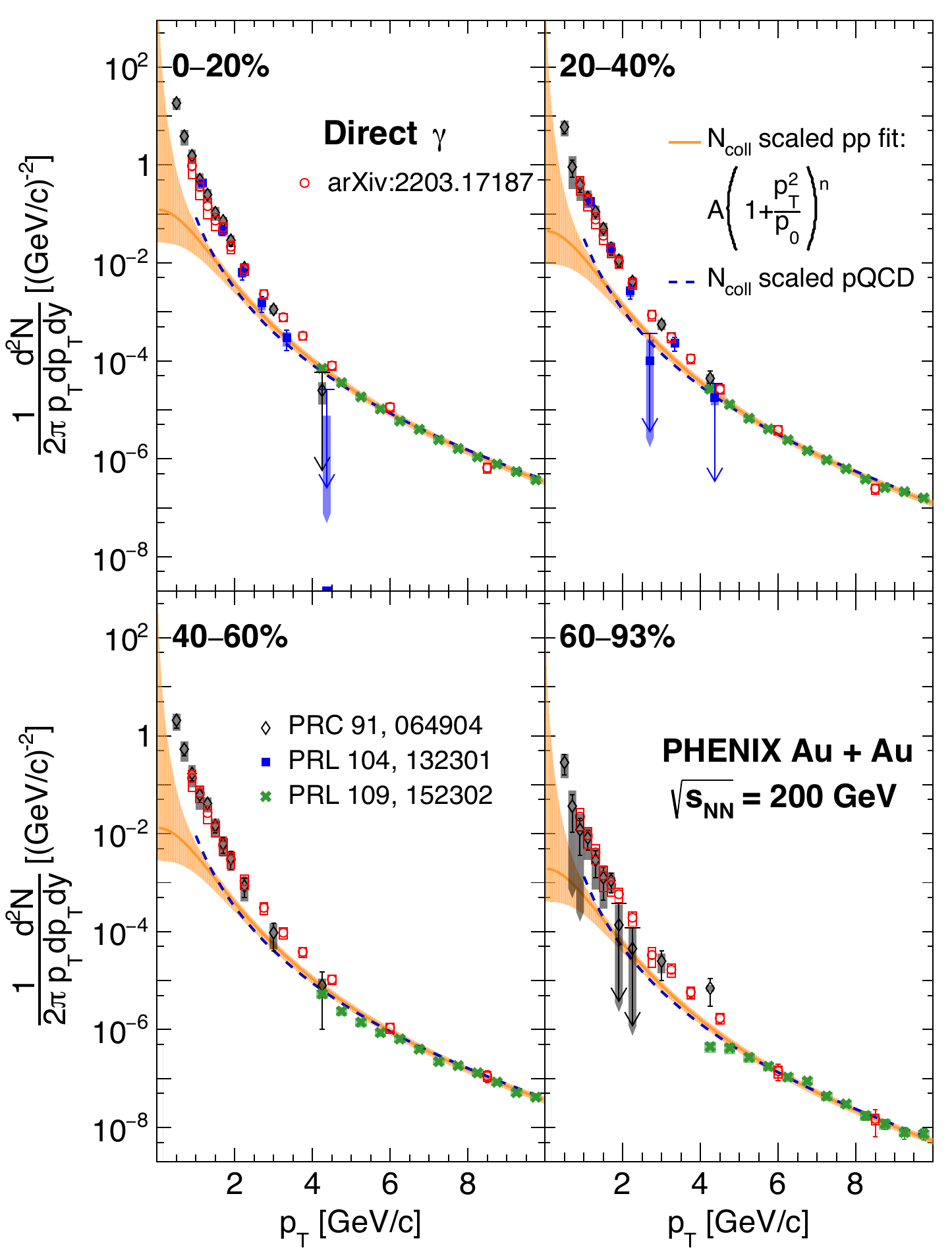}}
\caption{Invariant yield of direct photons as a function of $p_{T}$ for Au+Au at \snn{39} and \snn{64} (left) and for \snn{200} (right) for different collision centralities.}
\label{Fig:dp_spec}
\end{figure}

Photon conversions on the backplane of the Hadron Blind detector (HBD) were analyzed for Au+Au collisions 
recorded in 2010 at \snn{39} and \snn{62.4} and the corresponding direct photon spectra as a function of $p_{T}$ are shown in Fig.~\ref{Fig:dp_spec} (left) with the $T_{AA}$-scaled pQCD curve shown in solid line.

The substantial increase in the statistics for Au+Au collisions at \snn{200} allowed for a more detailed and differential measurement of direct photons. Instead of the HBD, which was removed, conversions in the layers of a new Silicon Vertex tracker, with a material budget of around 13\%, were analyzed. The spectrum for every 20\% collision centrality is shown in Fig.~\ref{Fig:dp_spec} (right). Remarkably, the results obtained from this new analysis exhibit excellent agreement with all the previous measurements conducted by PHENIX.

Having established the direct photon spectra, the subsequent step involves comprehending the dependence on collision centrality and investigating the shape of the spectrum.

\begin{figure}[htb]
\centerline{%
\includegraphics[height=4.0cm]{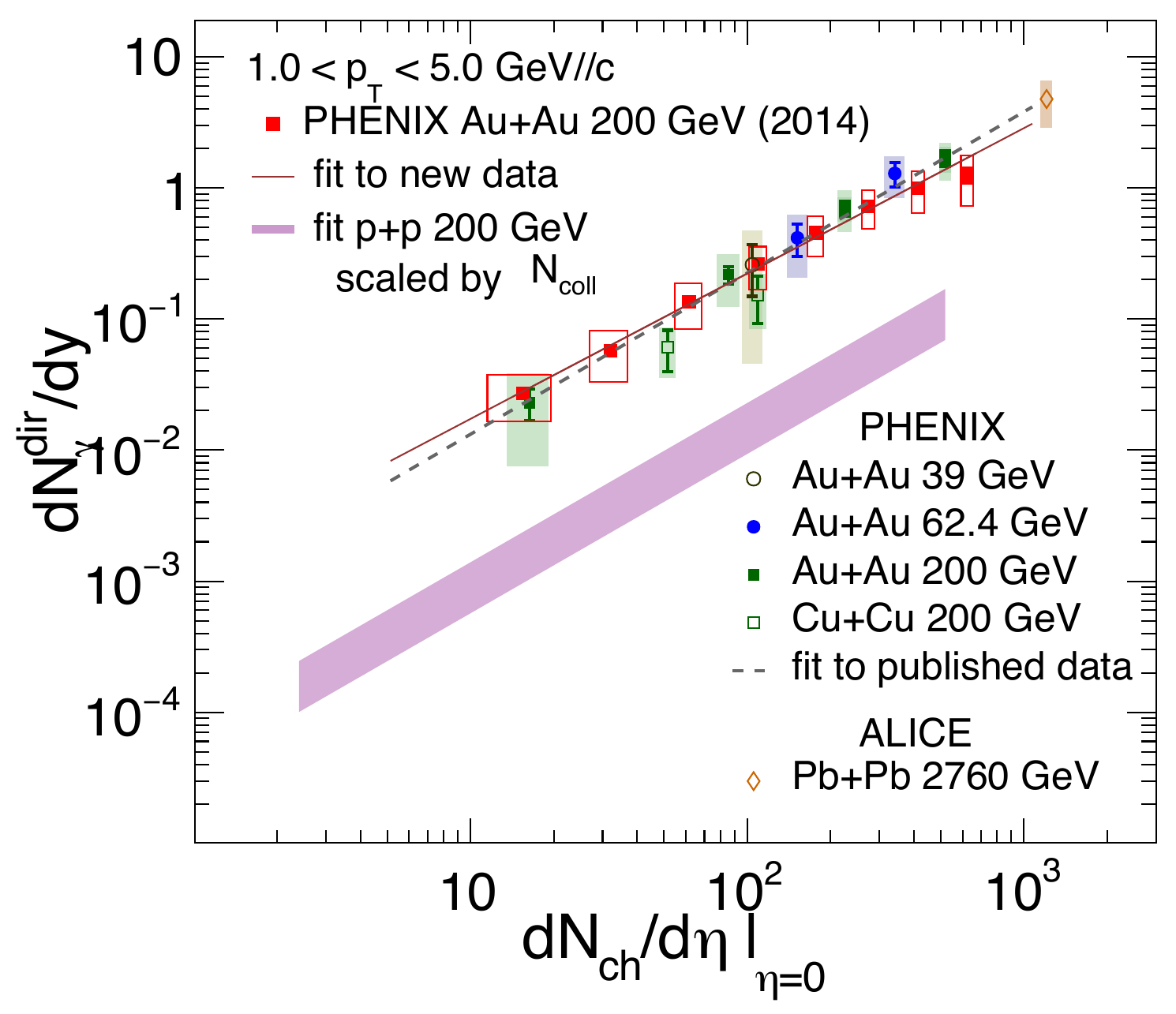} 
\hspace{2cm}
\includegraphics[height=4.0cm]{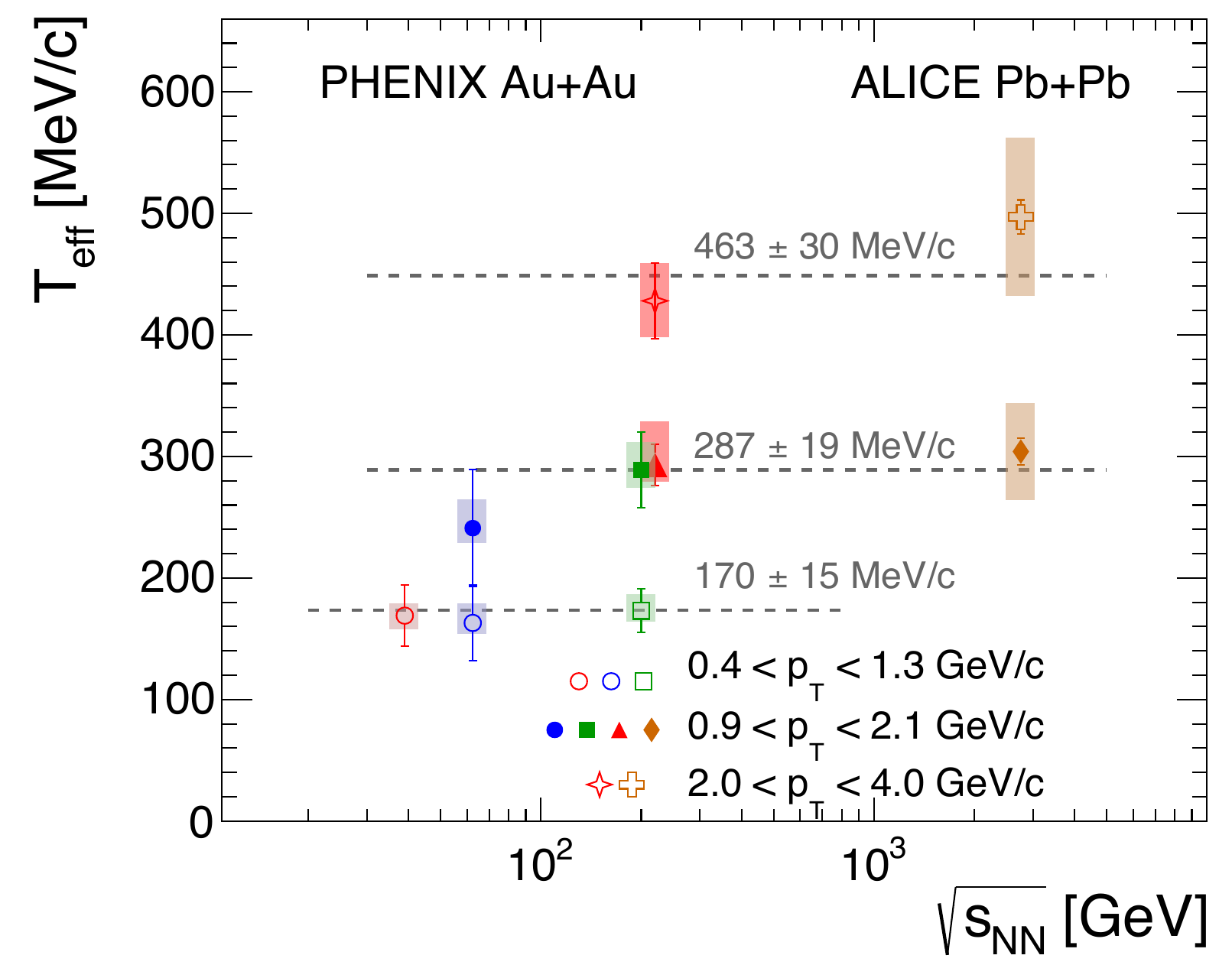}}
\caption{Integrated yield of direct photons as a function of system size (left) and the inverse slope of spectrum as a function of collision energy for different $p_{T}$ ranges (right).}
\label{Fig:dp}
\end{figure}

\textbf{Universal scaling}: To further investigate the centrality dependence, the integrated yields, in Figure \ref{Fig:dp} (left), are plotted as a function of charged particle multiplicity at midrapidity for various collision systems and energy spanning almost 2 orders of magnitude. Notably, a universal scaling behavior is observed across all $A+A$ collision systems, exhibiting a trend similar to that of scaled $p$+$p$ collisions but with yields approximately 10 times larger.

\textbf{Effective temperature}: The effective temperature can be determined by estimating the local inverse slope of the spectrum. To gain a deeper understanding of the similarities in the low-$p_{T}$ direct photon spectrum across different collision energies, the spectrum is fitted in various $p_{T}$ ranges. The resulting values of the effective temperature ($T_{\mathrm{eff}}$) extracted from these fits are displayed in Figure \ref{Fig:dp} (right). The consistency in extracted $T_{\mathrm{eff}}$ within different collision energies across various fit ranges implies that there are common sources responsible for direct photon production, regardless of the specific collision energy.

\section{Non-prompt direct photons}

\begin{figure}
\centerline{%
\includegraphics[height=5.45cm]{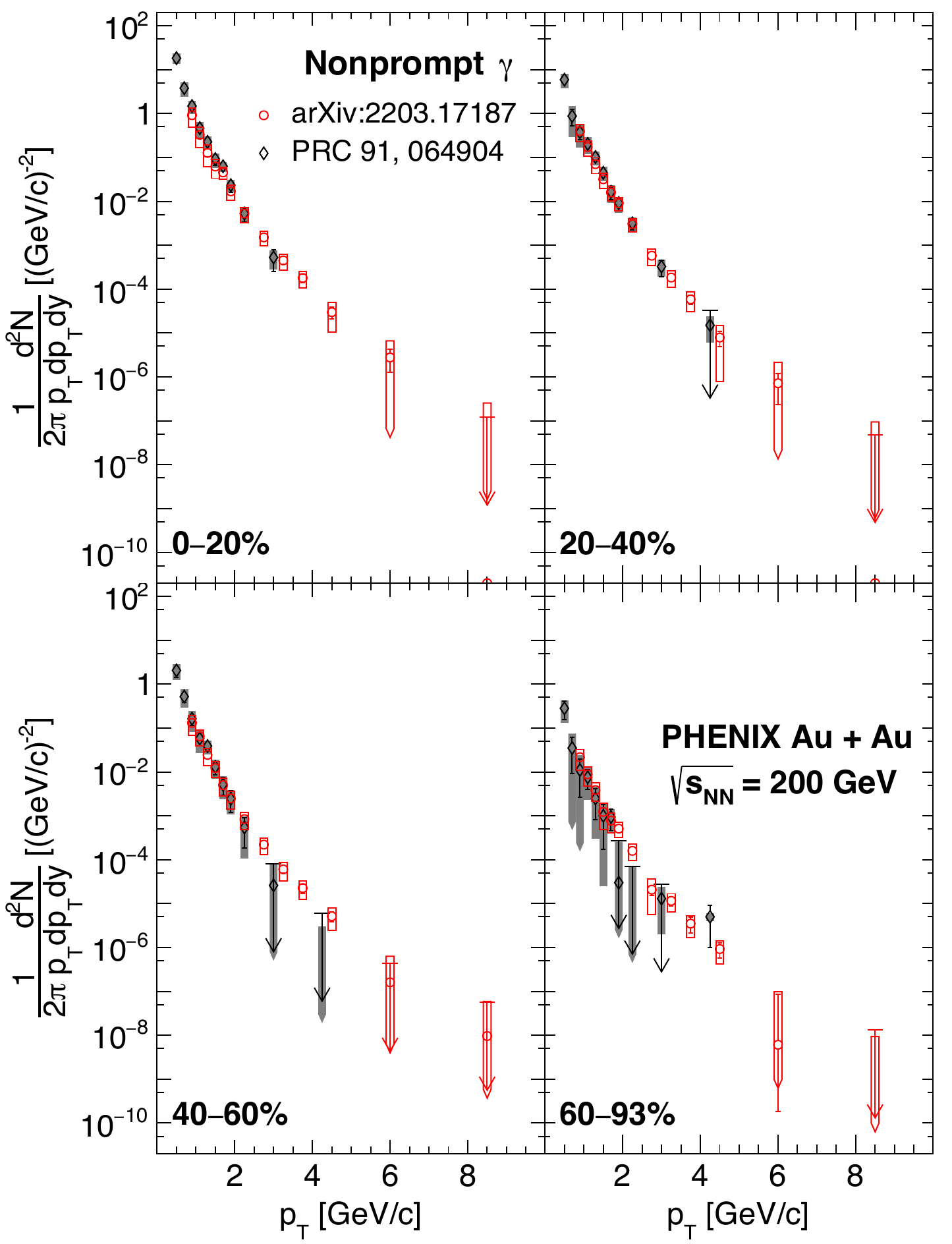}}
\caption{Invariant yield of non-prompt direct photons as a function of $p_{T}$ for Au+Au at \snn{200} for different collision centralities.}
\label{Fig:ndp_spec}
\end{figure}

Non-prompt direct photons are radiations that are emitted during the collision from the hot and expanding fireball and are estimated by subtracting the $N_{\mathrm{coll}}$ scaled p+p fit from the direct photon spectrum. The non-prompt direct photon spectra are shown for every 20\% collision centrality in Fig.~\ref{Fig:ndp_spec}. These measurements have significantly expanded their scope both in terms of $p_{T}$ coverage and centrality compared to previous publications.

\begin{figure}[htb]
\centerline{%
\includegraphics[height=4.0cm]{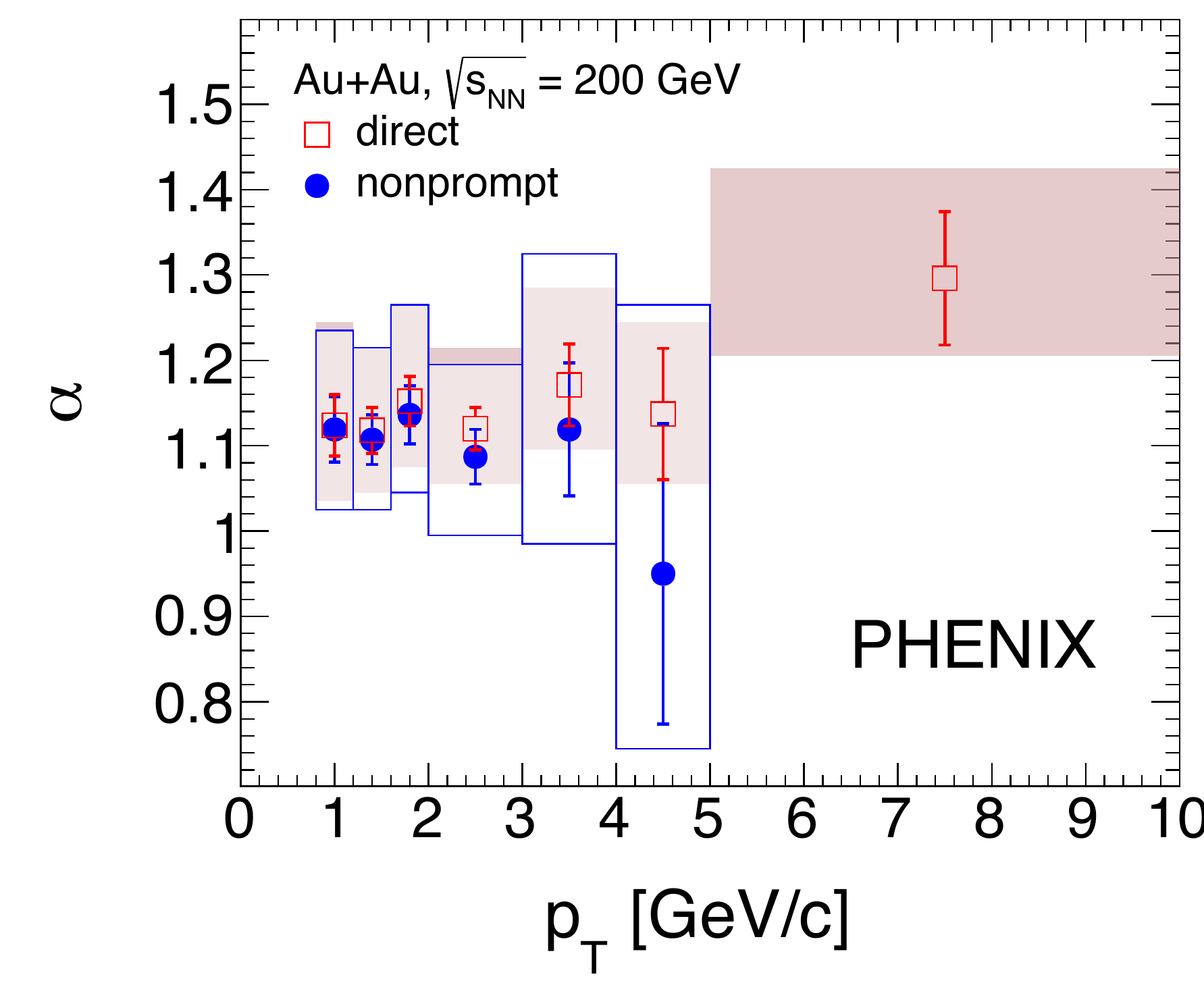}
\hspace{2cm}
\includegraphics[height=4.0cm]{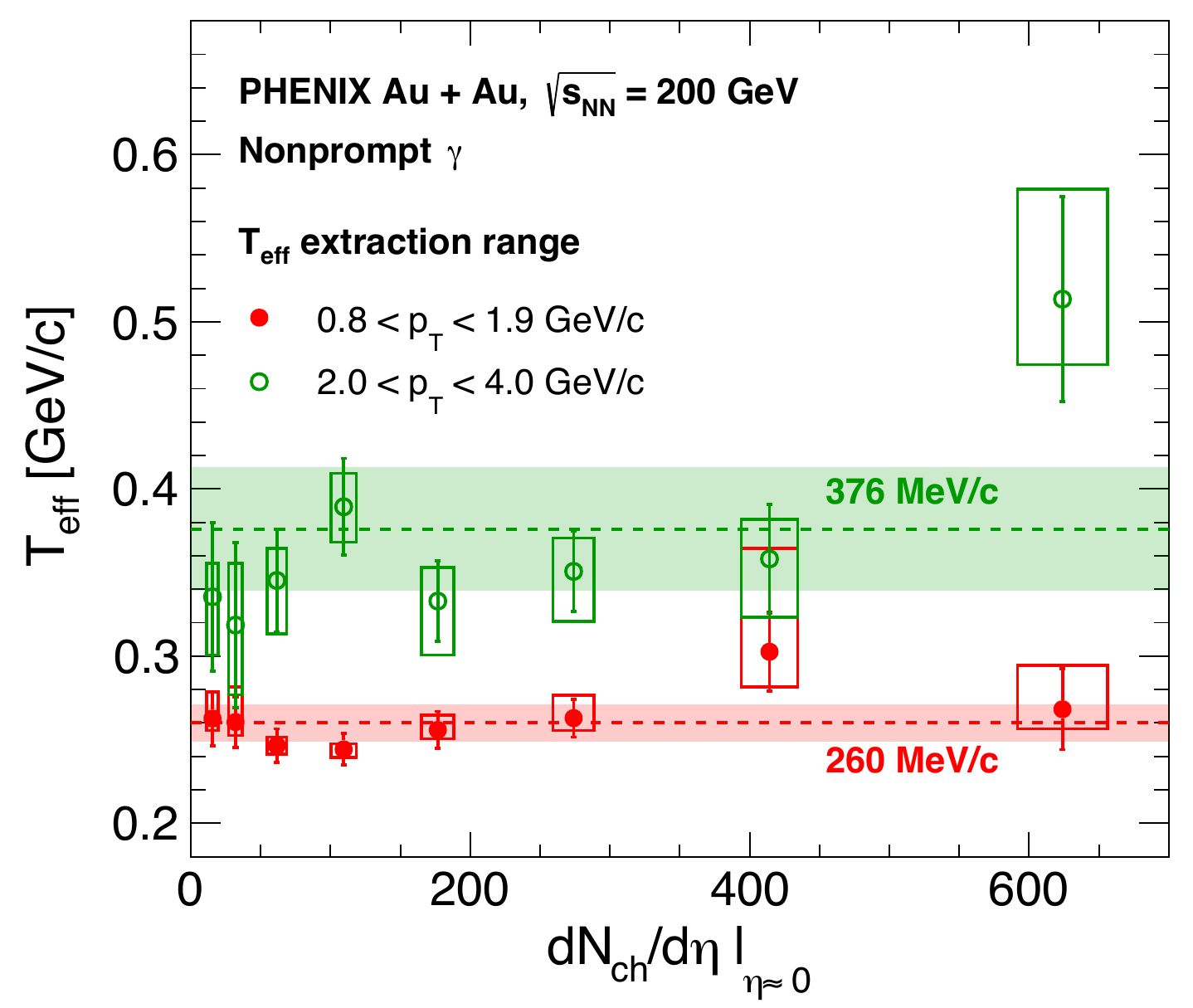}}
\caption{Scaling power $\alpha$, as a function of $p_{T}$ (left) and the inverse slope of non-prompt direct photon spectrum as a function of system size for different $p_{T}$ ranges (right).}
\label{Fig:npdp}
\end{figure}

\textbf{Universal scaling}: In order to investigate the centrality dependence, the scaling power, $\alpha$, is extracted by fitting the integrated yield as a function of charged particle multiplicity at midrapidity. Figure \ref{Fig:npdp} (left) displays the variation of $\alpha$ as a function of $p_{T}$ for six non-overlapping $p_{T}$ ranges for non-prompt direct and direct photon spectra. Below 3 GeV/$c$, the direct photon spectra are primarily influenced by non-prompt direct photon sources, resulting in similar values for $\alpha$. However, as $p_{T}$ increases, the $\alpha$ values begin to diverge, although it should be noted that the non-prompt direct photon spectra suffer from limited statistical precision. Experimental findings indicate that $\alpha$ remains relatively independent of $p_{T}$, contrary to the theoretical expectations that $\alpha$ increases as the system transitions to higher $p_T$ where the production is dominated by the QGP phase~\cite{Shen:2013vja}.

\textbf{Effective temperature}: The shape of the non-prompt direct photon spectrum is not described by a single exponential but rather has a continuously increasing inverse slope with $p_{T}$. To quantify this changing slope, the non-prompt direct photon spectra are fitted with exponentials in two distinct $p_{T}$ ranges, as depicted in Figure \ref{Fig:npdp} (right). The slopes are found to be consistent with a constant value and independent of the collision centrality. The average value of $T_{\mathrm{eff}}$ rises from 200 MeV to approximately 400 MeV within the $p_{T}$ range of 0.8 to 4 GeV/$c$. The variation in $T_{\mathrm{eff}}$ is not surprising, as the underlying spectra integrate the entire evolution of the expanding fireball, encompassing its earliest pre-equilibrium state, the Quark-Gluon Plasma (QGP) phase, the transition to the hadron gas phase, and subsequent expansion and cooling until hadrons cease interacting with each other. Consequently, contributions from the earliest phase are expected to dominate the spectra at higher $p_{T}$ values, which is consistent with the observation of an increasing $T_{\mathrm{eff}}$ with $p_{T}$.

\section{Comparisons to theory}
\begin{figure}
\centerline{%
\includegraphics[height=5.0cm]{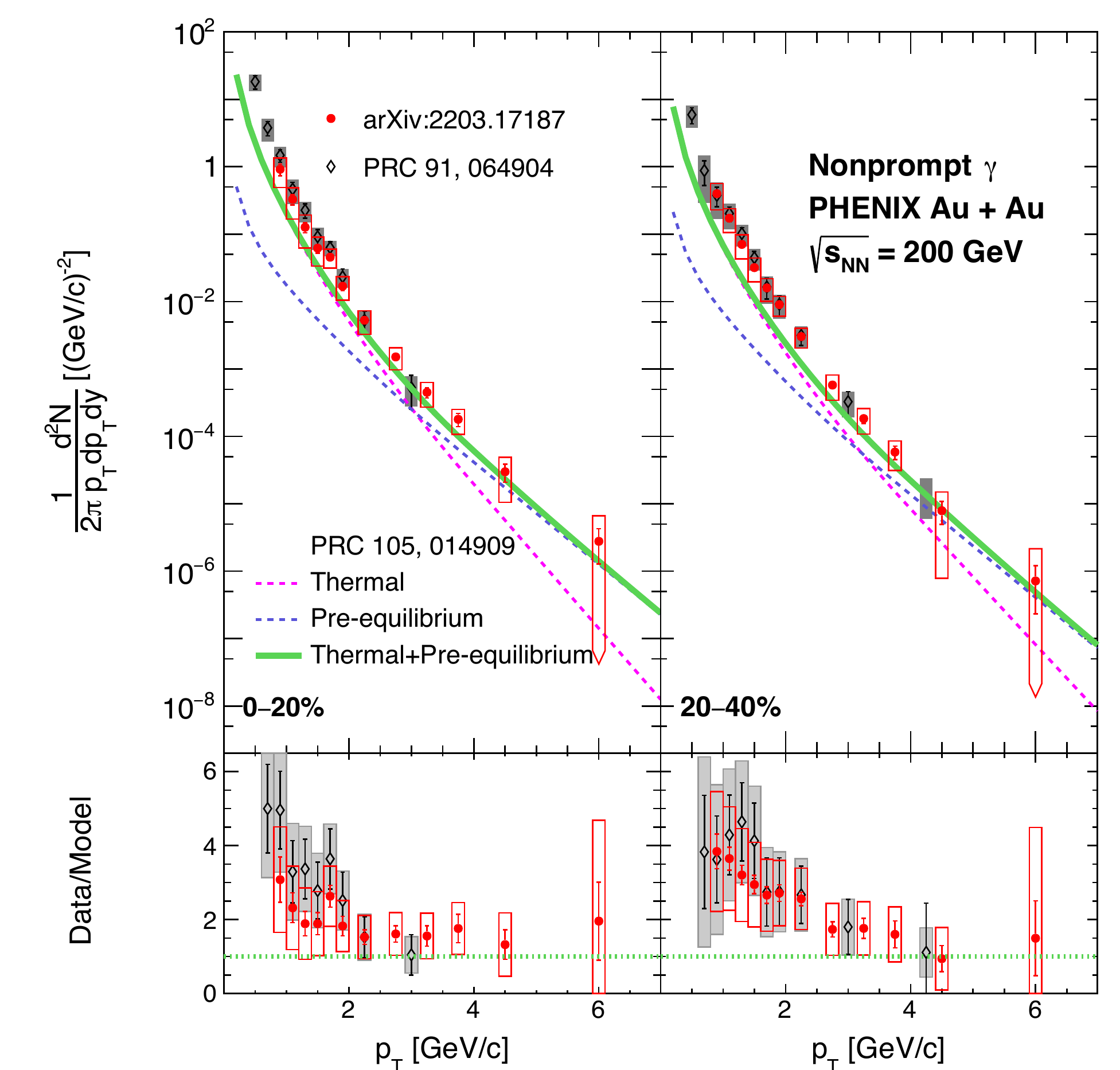} 
\hspace{2cm}
\includegraphics[height=4.0cm]{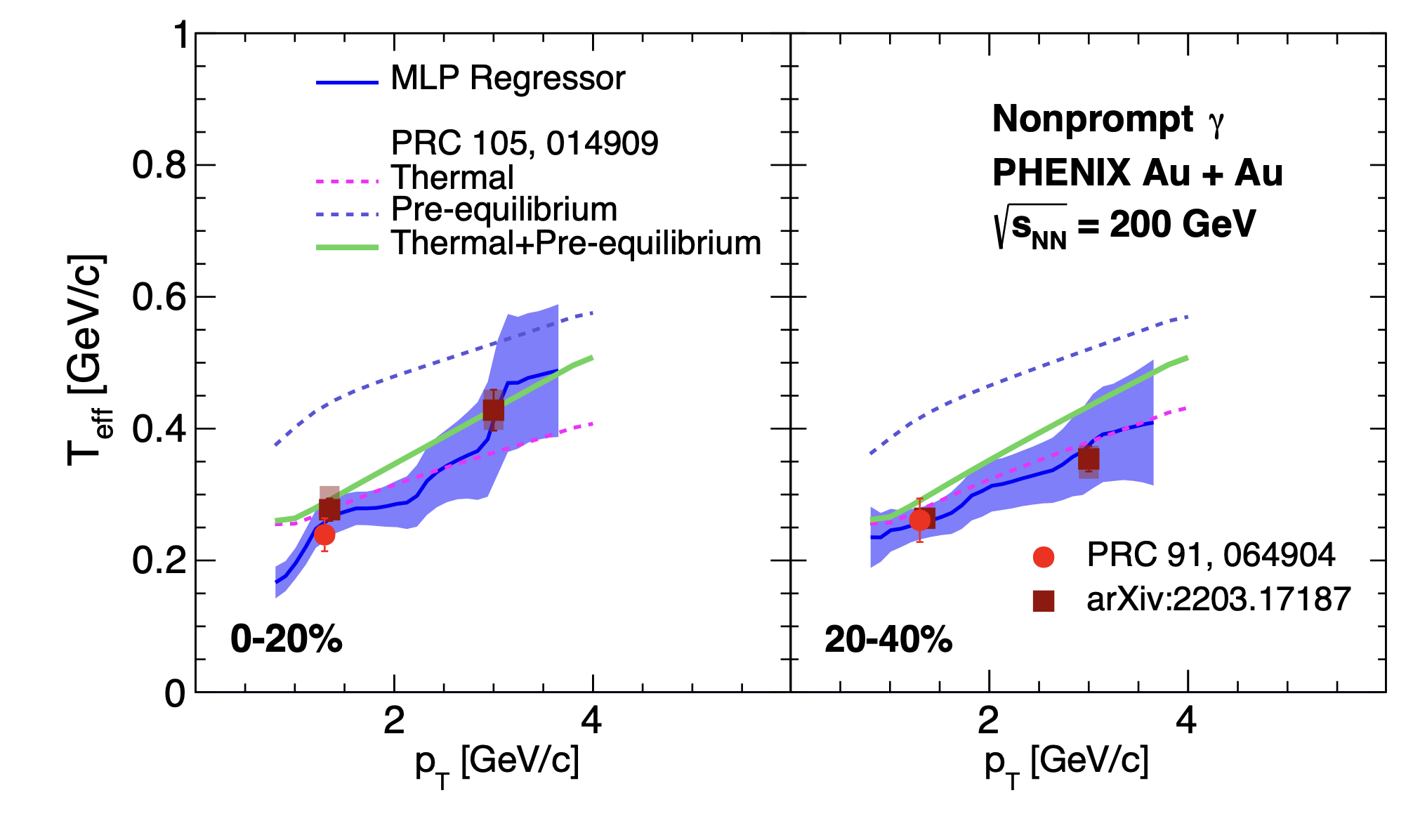}}
\caption{Comparison of the non-prompt direct photon spectrum for Au+Au at \snn{200} with theoretical calculations.}
\label{Fig:theory}
\end{figure}

In Figure \ref{Fig:theory} (left), the measured non-prompt direct photon spectra are compared to recent theoretical calculations that utilize a hybrid model, taking into account contributions from the pre-equilibrium state~\cite{Gale:2021emg}. The bottom panel of the figure presents the ratio of the measurements to the combined thermal and pre-equilibrium contributions predicted by the model. According to the calculations, the pre-equilibrium radiations are expected to become the dominant source of non-prompt direct photons above a $p_{T}$ of 3 GeV/$c$. While the shape of the spectra is well-reproduced by the model, the overall yield predicted by the model falls short, particularly below 2 GeV/$c$, where the measured yields appear to be approximately 2 to 3 times larger.

In order to further explore the shape of the non-prompt direct photon spectra, they are smoothened using a machine learning based regression algorithm called Multi Layer Perceptron on the PHENIX data~\cite{PHENIX:2022rsx}~\cite{PHENIX:2014nkk}. The inverse slope is extracted by numerical differentiation and is shown in Fig.~\ref{Fig:theory} (right). It can be argued that with increasing $p_{T}$, the contribution from the pre-equilibrium phase may be important.

\section{Direct photons in small systems}

\begin{figure}[htb]
\centerline{%
\includegraphics[height=4.8cm]{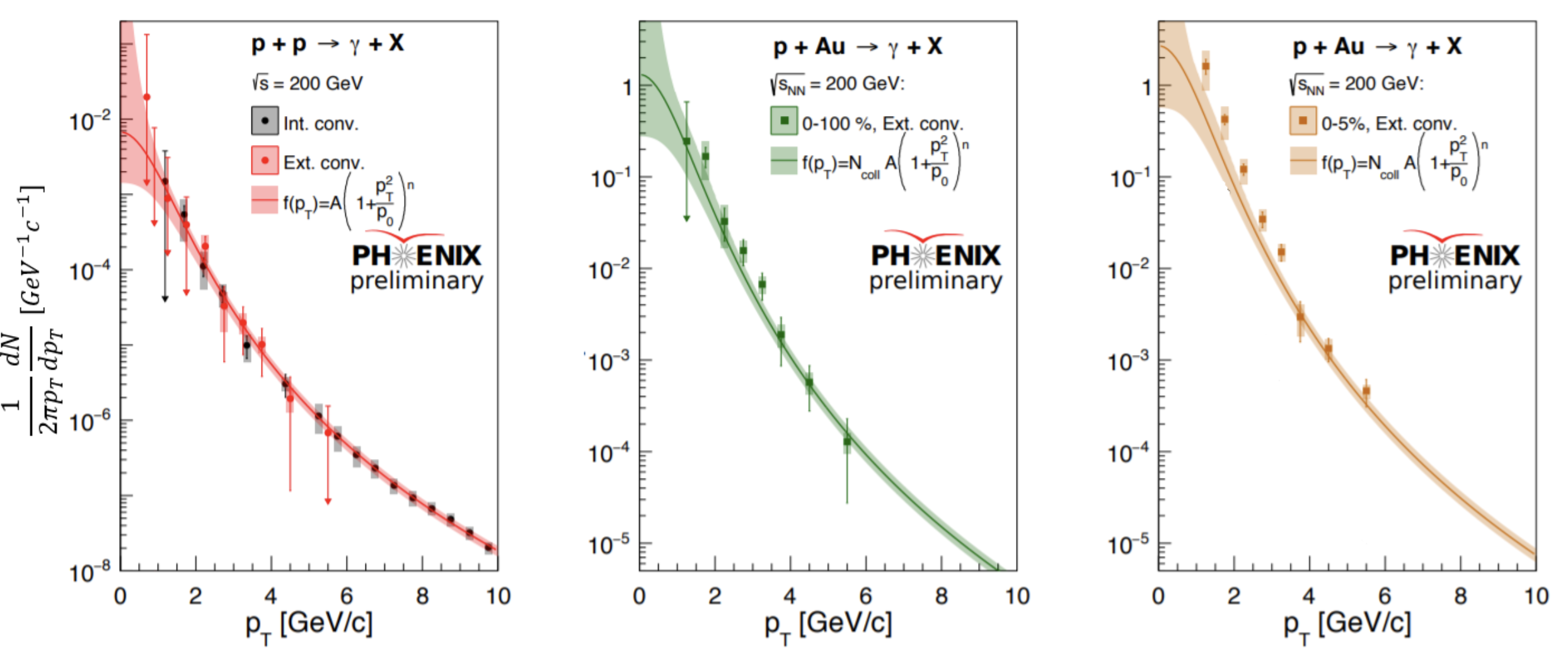} }
\caption{Direct photon spectrum for $p$+$p$ and $p+$+Au collisions at 200 GeV.}
\label{Fig:dp_pAu}
\end{figure}

For small systems, direct photons are measured for $p$+$p$ and $p$+Au collisions at 200 GeV using the data collected in 2015, utilizing external conversion in the layers of the Silicon Vertex detectors. In Fig.~\ref{Fig:dp_pAu}, the red data points obtained from external conversion in $p$+$p$ collisions aligns well with previous PHENIX measurements employing internal conversions. The solid line represents the $N_{\mathrm{coll}}$-scaled fit to the $p$+$p$ spectrum, which serves as a reference baseline. Notably, the lowest $p_T$ points for the most central $p$+Au collisions exhibit indications of an excess in the observed yields of direct photons.

\section{Summary}
In summary, the results of direct photon measurements in Au+Au collisions at energies of 39, 62.4, and 200 GeV are discussed. Additionally, a more detailed analysis focusing on non-prompt direct photons in high-statistics Au+Au collisions at 200 GeV is provided. A universal scaling behavior that remains consistent regardless of collision centrality, collision energy, or collision system, with respect to charged particle multiplicity at midrapidity is observed. Furthermore, the scaling power $\alpha$ shows an insignificant dependence on $p_{T}$ for both direct and non-prompt photons. Both direct photons and non-prompt direct photon spectra exhibit an increasing inverse slope as a function of $p_{T}$. Recent theoretical calculations including pre-equilibrium contributions seem to reduce the discrepancy between theory and observation. 

\bibliographystyle{JHEP.bst}
\bibliography{references}

\end{document}